\documentclass[10pt]{article}
\usepackage{amsmath}
\usepackage{amssymb}
\usepackage{natbib}

\begin{document}
\bibliographystyle{plainnat}
\setcitestyle{numbers,square}

\title{Gravitational quantization of exoplanet orbits in                  \\
                   HD~40307, $\mu$~Ara, Kepler-26, Kepler-62, and Kepler-275: 
                   Comparing predicted orbits}

\author{Vassilis S. Geroyannis \\
        Department of Physics, University of Patras, Greece  \\  
        vgeroyan@upatras.gr}

\maketitle

\begin{abstract}
According to the so-called ``global polytropic model'', hydrostatic equilibrium for a planetary system leads to the Lane--Emden differential equation. Solving this equation in the complex plane, we obtain polytropic spherical shells defined by succesive roots of the real part $\mathrm{Re}(\theta)$ of the Lane--Emden function $\theta$. Such shells provide hosting orbits for the members of a planetary system. Within the framework of the global polytropic model, we study the exoplanet systems HD~40307, $\mu$~Ara, Kepler-26, Kepler-62, and Kepler-275. We give emphasis on comparing our results with observations, and with orbit predictions derived from the ``generalized Titius--Bode relation''.   
\end{abstract}

\section{Introduction}
\label{intro}
Gravitational quantization of orbits in systems of planets or satellites has been considered in three recent investigations (\citep{GVD14}, \citep{G14}, \citep{G14b}). In the framework of classical mechanics, such systems are assumed to be in hydrostatic equilibrium, governed by the Lane--Emden differential equation. We solve this equation in the complex plane applying the so-called ``complex plane strategy'' \citep{GKAR14}. The resulting solution is the complex Lane--Emden function $\theta$. Polytropic spherical shells defined by succesive roots of the real part $\mathrm{Re}(\theta)$ seem to provide hosting orbits for planets or satellites. In fact, there is only one parameter to be adjusted for a particular polytropic configuration defined by $\theta$: the polytropic index $n$ of the central body. 
  
Alternative studies regarding quantization of orbits of planets or satellites are cited and discussed in \citep{GVD14} and \citep{G14}.

In the present paper, we emphasize on comparing our results with: (i) corresponding observations, and (ii) orbit predictions of undetected exoplanets, calculated by the so-called ``generalized Titius--Bode relation'' (\cite{BL13}, Secs.~1--3 and references therein; the method is analyzed in Sec.~4). Table~2 of \citep{BL13} shows orbit predictions for 29 exoplanet systems; in this table, there are both ``interpolated predictions'' concerning undetected planets with distances less than that of the outermost planet, and ``extrapolated predictions'' regarding  undetected planet(s) being next to the outermost detected one(s). For the purpose of examining, as best we can, the issue (ii) above, we focus on 5 exoplanet systems with the larger numbers of predictions: HD~40307 (4+1 predictions), $\mu$~Ara (4+1 predictions), Kepler-26 (5+1 predictions), Kepler-62 (6+1 predictions), and Kepler-275 (6+1 predictions).

\section{Polytropic Models Simulating Host Stars, and Polytropic Shells Hosting Exoplanets}
\label{gpm}
Preliminary concepts for the global polytropic model can be found in \citep{GVD14} (Sec.~3).
The complex-plane strategy and the complex Lane--Emden function $\theta$ are analysed in \citep{GKAR14} (Sec.~3.1). We will use hereafter the definitions and symbols adopted in \cite{GVD14}. 

The real part $\bar{\theta}(\xi)$ of $\theta(\xi)$ has a first root $\xi_1 = \bar{\xi_1} + i \, \breve{\xi_0}$, which is the radius of the central body. Next to $\xi_1$, there is a second root $\xi_2 = \bar{\xi_2} + i \, \breve{\xi_0}$ with $\bar{\xi_2} > \bar{\xi_1}$, a third root $\xi_3 = \bar{\xi_3} + i \, \breve{\xi_0}$ with $\bar{\xi_3} > \bar{\xi_2}$, etc. The polytropic spherical shells $S_2$, $S_3$, \dots, are defined by the pairs $(\bar{\xi}_1, \, \bar{\xi}_2)$, $(\bar{\xi}_2, \, \bar{\xi}_3)$, \dots, respectively.
A shell $S_j$ can be considered as a place appropriate for hosting a planet. For this purpose, the most proper orbit radius $\alpha_j \in [\bar{\xi}_{j-1},\,\bar{\xi_j}]$ is that at which $|\bar{\theta}|$ takes its maximum value inside $S_j$, 
\begin{equation}
\mathrm{max}|\bar{\theta}[S_j]| = |\bar{\theta}(\alpha_j + i \, \breve{\xi}_0)|.
\label{maxth} 
\end{equation}
In the case of two or three planets hosted inside the same shell $S_j$, there are two further proper orbits with radii $\alpha_\mathrm{Lj}$ and $\alpha_\mathrm{Rj}$, such that $\alpha_\mathrm{Lj} < \alpha_j < \alpha_\mathrm{Rj}$, at which $|\bar{\theta}|$ becomes equal to its average value inside $S_j$,
\begin{equation}
\mathrm{avg}|\bar{\theta}[S_j]| = |\bar{\theta}(\alpha_\mathrm{Lj} + i \, \breve{\xi}_0)|
                                = |\bar{\theta}(\alpha_\mathrm{Rj} + i \, \breve{\xi}_0)|.
\label{avgth}   
\end{equation}
So, planets inside $S_j$ can be hosted on orbits with radii  $\alpha_\mathrm{Lj}$, $\alpha_\mathrm{j}$, and $\alpha_\mathrm{Rj}$.   

In \citep{G14} (Sec.~2), we have developed an algorithm for computing the optimum polytropic index $n_\mathrm{opt}$ of a star, which hosts a number of planets. This algorithm, called A[n], can be applied to $N_\mathrm{P}$ members $P_1,\,P_2,\,\dots,\,P_{N_\mathrm{P}}$ of a system with $N_\mathrm{P}$ prescribed distances $A_{1} < A_{2} < \dots < A_{{N_\mathrm{P}}}$ from the central body.
 
The host stars of the exoplanet systems HD~40307 (\citep{PT10}, \citep{TAG12}, \citep{YJY14}), $\mu$~Ara (\citep{GKM05}, \citep{PCM07}, \citep{WTO10}), Kepler-26 (KOI-250;  \citep{SFF12}, \citep{D12}), Kepler-62 (KOI-701; \citep{BAF13}, \citep{K14}, \citep{BRL14}), and Kepler-275 (KOI-1198; \citep{RBM14}) are Sun-like stars. We expect therefore that the optimum values of the polytropic index $n$ for modeling such stars are about $n \sim 3$ (\citep{Hor04}, Sec.~6.1 and references therein; \citep{GVD14}, Sec~3 and references therein). 
Accordingly, we apply the algorithm A[n] to an array $\{n_i\}$ with elements
\begin{equation}
n_i = 2.500 + 0.001 \, (i-1), \qquad i = 1, \, 2, \, \dots, \, 1001.
\label{Nn-now}
\end{equation} 
The 1001 complex IVPs, counted in this equation, are solved by the Fortran code \texttt{DCRKF54} \cite{GV12}, which is a Runge--Kutta--Fehlberg code of fourth and fifth order modified for solving complex IVPs of high complexity in their ODEs, along contours prescribed as continuous chains of straight-line segments (for details on the usage of \texttt{DCRKF54}, see \citep{GVD14}, Sec.~4; see also \citep{G14b}, Sec.~3). 

We will hereafter quote the real parts of the complex orbit radii, since these quantities have the physical interest in this study. For simplicity, we will drop overbars denoting these real parts.

\section{Numerical Results and Discussion}
\label{results}
All astrophysical data used for comparing our results with exoplanet observations are from the site http://exoplanet.eu. 

In the following tables, the first root $\xi_1$ of $\theta$, coinciding with the radius of the host star, is expressed in both ``classical polytropic units'' (cpu) --- in such units, the length unit is equal to the polytropic parameter $\alpha$ (\citep{GVD14}, Eq.~(3b)) --- and solar radii $R_\odot$. All other orbit radii are expressed in AU.

\subsection{The HD~40307 system}
Results for the HD~40307 system are shown in Table~\ref{hd40307}. For this system, the minimum sum of absolute percent errors is 
\begin{equation}
\begin{aligned}
\Delta_\mathrm{min} 
\biggl( n_\mathrm{opt}(\mathrm{HD\,40307}) & =
        2.531; \, q_\mathrm{b} = 5, \, q_\mathrm{c} = 6, \biggr. \\
      & \biggl.   q_\mathrm{d} = 8, \, q_\mathrm{e} = 9, \, q_\mathrm{f} = 10, \,
                  q_\mathrm{g} = 15 
\biggr) \simeq 37.9.
\label{DminJnow}
\end{aligned}
\end{equation}
Smaller error is that for g's distance, $\simeq 2.2\%$, and larger one is that for f's distance, $\simeq 16\%$. For the most massive planet, d, the error is $\simeq 3.3\%$. 
The average error for the computed orbit radii of the 6 planets in HD~40307 is $\simeq 6.3\%$.

Regarding the large error quoted for f's distance, an interesting conjecture is to associate this distance with the mean-density orbit $\alpha_\mathrm{R10} = 0.2274$~AU, provided that the maximum-density orbit $\alpha_{10}$ is already occupied by another planet not yet observed. In view of this conjecture, the error for f's distance drops to $\simeq 7.9\%$, the minimum sum of absolute percent errors drops to $\simeq 30\%$, and the average error is reduced to $\simeq 5.1\%$.  

Table~2 of \citep{BL13} includes 5 predictions of planet orbits for the HD~40307 system. To compare these predictions with orbit radii computed by A[n], we present in Table \ref{hd40307IE} all internal shells, i.e. shells located before the outermost occupied shell No~15, which seem to be unoccupied according to the up-to-now observations. Comparison is also made for the external shell No~17, within which an extrapolated planet prediction seems to exist according to \citep{BL13}. The sum of absolute percent differences between corresponding estimates is $\simeq 32\%$, giving an average difference $\simeq 6.3\%$. 

The larger difference, $\simeq 12\%$, is found for the shell No~12. On this issue, it seems interesting to apply again the conjecture discussed above. In particular, the prediction $P_3 = 0.33$~AU can be associated with the orbit $\alpha_\mathrm{L12} = 0.3496$~AU, leading to a smaller difference $\simeq 5.9\%$. However, if, eventually, the average-density orbit $\alpha_\mathrm{L12} \simeq 0.35$~AU hosts a planet, then the maximum-density orbit $\alpha_{12} \simeq 0.37$~AU should already host a planet.

\subsection{The $\mu$~Ara system}
As seen in Table~\ref{mAr}, the optimum case for the $\mu$~Ara system gives minimum sum of absolute percent errors 
\begin{equation}
\Delta_\mathrm{min} \biggl( n_\mathrm{opt}(\mathrm{\mu~Ara}) = 2.554; \, q_\mathrm{c} = 5, \, q_\mathrm{d} = 14, \, q_\mathrm{b} = 18, \, q_\mathrm{e} = 31 \biggr) \simeq 5.39.
\label{DminS}
\end{equation}
Smaller error is that for e's distance, $\simeq 0.06\%$, which is also the most massive planet; while larger error is that for c's distance, $\simeq 3.9\%$.
The average error for the computed distances of the 4 planets in $\mu$~Ara is $\simeq 1.3\%$.

Regarding predictions of planetary orbits, Table~2 of \citep{BL13} shows 5 such cases for the $\mu$~Ara system. Table \ref{mArIE} presents shells providing respective hosting orbits computed by A[n]. The sum of  absolute percent differences of associated estimates is $\simeq 27\%$, which leads to an average difference $\simeq 5.4\%$. 

The larger discrepancy, $\simeq 15\%$, appears in the shell No~8. Alternatively, we can conjecture, as above, that the prediction $P_2 = 0.29$~AU is associated with the orbit $\alpha_\mathrm{R8} = 0.2714$~AU, deriving a smaller difference  $\simeq 6.4\%$. Accordingly, if, eventually, the average-density orbit $\alpha_\mathrm{R8} \simeq 0.27$~AU hosts a planet, then the maximum-density orbit $\alpha_{8} \simeq 0.25$~AU should already host a planet.

\subsection{The Kepler-26 (KOI-250) system}
For the Kepler-26 system, Table~\ref{k26} reveals the optimum case     
\begin{equation}
\Delta_\mathrm{min} \biggl( n_\mathrm{opt}(\mathrm{Kepler\!-\!26}) = 
2.531; \, q_\mathrm{d}=5, \,
          q_\mathrm{b}=7, \, q_\mathrm{c}=8, \, q_\mathrm{e}=11 \biggr) \simeq 18.57.
\label{DminU}
\end{equation}
Here, smaller error is that for b's distance, $\simeq 2.9\%$, while larger one is that for e's distance, $\simeq 7.2\%$.
The average error in the computed distances of the 4 planets in Kepler-26 is $\simeq 4.6\%$.

Table~2 in \citep{BL13} shows 6 predictions of planet orbits for the Kepler-26 system. In Table~\ref{k26IE}, we present all unoccupied shells according to up-to-now observations, located before the outermost occupied shell No~11. The external shell No~12, for which an extrapolated planet prediction is made in \citep{BL13}, is also included in this table.
The sum of absolute percent differences of respective estimates is $\simeq 39\%$, leading to an average difference $\simeq 6.5\%$. 

The larger discrepancy, $\simeq 16\%$, arises in the shell No~9. This shell hosts the orbit $\alpha_\mathrm{L9} \simeq 0.16$~AU associated with the prediction $P_4 = 0.14$~AU, and the orbit $\alpha_9 \simeq 0.17$~AU associated with the prediction $P_5 = 0.17$~AU. The latter almost coincides with its counterpart, while the former is in large discrepancy with its counterpart (that quoted above). Since this discrepancy is $\sim 3$ times the average difference, we adjudge this issue in favour of the orbit $\alpha_\mathrm{L9} \simeq 0.16$~AU computed by the algorithm A[n], and we thus adopt this orbit in the place of the prediction $P_4 = 0.14$~AU obtained by the Titius--Bode relation. 

The second larger discrepancy, $\simeq 13\%$, appears in the shell No~12. We can again  conjecture that the prediction $P_\mathrm{6} = 0.27$~AU is associated with the orbit  $\alpha_\mathrm{L12} = 0.2881$~AU, giving an appreciably smaller difference $\simeq 6.7\%$. In view of such a conjecture, if, eventually, the orbit $\alpha_\mathrm{L12} \simeq 0.29$~AU is occupied by a planet, then the orbit $\alpha_{12} \simeq 0.30$~AU should be already occupied by a planet.

\subsection{The Kepler-62 (KOI-701) system}
For the Kepler-62 system, Table~\ref{k62} gives the optimum case   
\begin{equation}
\begin{aligned}
\Delta_\mathrm{min} \biggl( n_\mathrm{opt}(\mathrm{Kepler\!-\!62}) & = 
2.773; \, q_\mathrm{b}=5, \biggr. \\ 
& \biggl. 
  q_\mathrm{c}=6, \, \, q_\mathrm{d}=7, \, q_\mathrm{e}=11, \, q_\mathrm{f}=14   \biggr) \simeq 20.36.
\label{DminK20}
\end{aligned}
\end{equation}
Smaller error is that for b's distance, $\simeq 0.08\%$, while larger error is that for c's distance, $\simeq 14\%$. The error for the most massive planet, e, is $\simeq 0.42\%$.
The average error in the computed distances of the 5 planets in Kepler-62 is $\simeq 4.1\%$.

Regarding comparisons with the 7 predicted orbits of the Kepler-62 system in Table~2 of \citep{BL13}, Table \ref{k62IE} reveals that the sum of absolute percent differences of respective estimates is $\simeq 44\%$, which gives an average difference $\simeq 6.2\%$. The larger discrepancy, $\simeq 14\%$, arises for the orbit $\alpha_{13}$ associated with the prediction $P_6 = 0.55$~AU. Instead, we can again conjecture that the prediction $P_6$ is associated with the orbit  $\alpha_\mathrm{L13} = 0.5870$~AU, resulting in a smaller difference $\simeq 6.7\%$. Accordingly, if, eventually, the orbit $\alpha_\mathrm{L13} \simeq 0.59$~AU is occupied by a planet, then the orbit $\alpha_{13} \simeq 0.63$~AU should be also occupied by a planet. In fact, we have tacitly invoked this conjecture for the prediction $P_5 = 0.33$~AU, which has been associated with the orbit $\alpha_\mathrm{R10}$ instead of the orbit $\alpha_{10} = 0.2992$~AU (taking the latter in the place of the former, we find a difference $\simeq 9.3\%$, a total difference $\simeq 53\%$, and an average difference $\simeq 7.5\%$).

It is worth noting here that the shells No~5 and No~7, hosting the planets b and d on their maximum-density orbits $\alpha_5$ and $\alpha_7$, do also provide the average-density orbits $\alpha_\mathrm{R5}$ and $\alpha_\mathrm{R7}$, which are associated with the predictions $P_1$ and $P_2$, respectively. We can consider this issue as the ``complementary case'' of the conjecture discussed above. In particular, if we assume for a while that the planets b and d have not been observed yet, then our eventual conclusion that $\alpha_\mathrm{R5}$ corresponds to $P_1$ and $\alpha_\mathrm{R7}$ corresponds to $P_2$, does also lead to the conclusion that $\alpha_5$ and $\alpha_7$ should be already occupied by planets.

\subsection{The Kepler-275 (KOI-1198) system}
For the exoplanet system of Kepler-275, Table~\ref{k275} gives the optimum case 
\begin{equation}
\Delta_\mathrm{min} \biggl( n_\mathrm{opt}(\mathrm{Kepler\!-\!275}) = 
2.837; \, q_\mathrm{b}=4, \, q_\mathrm{c}=5, \, q_\mathrm{d}=6
                    \biggr) \simeq 10.5.
\label{DminK90}
\end{equation}
Smaller error is that for b's distance, $\simeq 0.42\%$, while larger one is that for c's distance, $\simeq 5.4\%$, which is also the most massive planet.
The average error in the computed distances of the 3 planets in Kepler-275 is $\sim 3.5\%$.

Table~2 of \citep{BL13} gives 7 predictions for the Kepler-275 system. As seen in Table~\ref{k275IE}, the sum of absolute percent differences of respective estimates is $\simeq 58\%$, giving an average difference $\simeq 8.3\%$. We now focus our attention on two cases with large discrepancies. The first case has to do with the prediction $P_1 = 0.03$~AU and its associated orbit $\alpha_\mathrm{R2} \simeq 0.022$~AU. The resulting large discrepancy $\simeq 25\%$ ($\sim 3$ times the average discrepancy) leads us to adjudge this issue in favour of the orbit $\alpha_\mathrm{R2}$. Second, likewise, the large discrepancy $\simeq 15\%$ ($\sim 2$ times the average discrepancy) between the prediction $P_6 = 0.17$~AU and its associated orbit $\alpha_\mathrm{R5} \simeq 0.14$~AU turns to be in favour of the latter. Thus we adopt the orbits $\alpha_\mathrm{R2}$ and $\alpha_\mathrm{R5}$ computed by the algorithm A[n] in the place of the predictions $P_1$ and $P_6$ obtained by the Titius--Bode relation. Since however $\alpha_\mathrm{R2}$ is an average-density orbit, adopting it as a planet's orbit prediction, presumes that the maximum-density orbit $\alpha_2$ should have priority for hosting a planet. It is worth noting here that, for the similar case of $\alpha_\mathrm{R5}$, the maximum-density orbit $\alpha_5$ is already occupied by the planet c (Table~\ref{k275}).   

Finally, it is of particular interest the fact that, in order to associate three orbits with the predictions $P_2 = 0.04$~AU, $P_3 = 0.05$~AU, and $P_4 = 0.06$, we need to employ all three available orbits in the shell No~3, i.e. $\alpha_\mathrm{L3} \simeq 0.038$~AU, $\alpha_3 \simeq 0.051$~AU, and $\alpha_\mathrm{R3} \simeq 0.061$~AU, respectively. Accordingly, any further prediction in the interval $[\sim 0.03 \, \mathrm{AU}, \, \sim 0.07 \, \mathrm{AU}]$ should be rejected.

\subsection{Further discussion on the HD~40307 and Kepler-26 systems}       
Tables~\ref{hd40307} and \ref{k26} show that HD~40307 and Kepler-26 both have optimum polytropic index $n_\mathrm{opt} = 2.531$. This means that all respective orbit radii are equal when expressed in polytropic units. Equivalently, all respective ratios of orbit radii are equal; for example, 
\begin{equation}
\left( \frac{\xi_5}{\xi_8} \right)_\mathrm{HD \, 40307} =
\left( \frac{\xi_5}{\xi_8} \right)_\mathrm{Kepler-26} = 0.3925,                                                        
\end{equation}  
\begin{equation}
\left( \frac{\alpha_5}{\alpha_8} \right)_\mathrm{HD \, 40307} =
\left( \frac{\alpha_5}{\alpha_8} \right)_\mathrm{Kepler-26} = 0.3593,
\end{equation}
etc. 

Next, the orbits $\alpha_5$ and $\alpha_8$ are occupied in both systems. In addition, the orbit $\alpha_\mathrm{R5}$ is found to be in satisfactory agreement with respective predictions of \citep{BL13} in both systems.

Furthermore, the orbits $\alpha_\mathrm{L6}$, $\alpha_6$, $\alpha_\mathrm{L9}$, and $\alpha_9$ in the Kepler-26 system are found to be in satisfactory agreement with respective predictions of \cite{BL13}; while the orbits $\alpha_6$ and $\alpha_9$ in the HD~40307 system are occupied. This means in turn that the average-density orbits $\alpha_\mathrm{L6}$ and $\alpha_\mathrm{L9}$ could be also predictions for the HD~40307 system, but they are ``missing'' from Table~2 of \citep{BL13}. 

Relevant worth noting cases are: (i) the orbit $\alpha_7$ occupied in the Kepler-26 system, and being in satisfactory agreement with respective prediction of \citep{BL13} for the HD~40307 system; and (ii) the orbit $\alpha_{11}$ occupied in the Kepler-26 system, and being empty in \citep{BL13}; so, the latter orbit could be a prediction for the HD~40307 system, but it is ``missing'' from Table~2 of \citep{BL13}. 

Finally, the following cases seem to be also ``missing predictions'' from Table~2 of \citep{BL13}: (i) the orbit $\alpha_{10}$ in Kepler-26, while this orbit is occupied in HD~40307; (ii) the orbit $\alpha_\mathrm{13}$ in Kepler-26, while this orbit is quoted as prediction for HD~40307 in \citep{BL13}; (iii) the orbit $\alpha_{15}$ in Kepler-26, while this orbit is occupied in HD~40307; and (iv) the orbit $\alpha_\mathrm{17}$ in Kepler-26, while this orbit is quoted as prediction for HD~40307 in \citep{BL13}.

\section{Conclusions}
We have examined 5 exoplanetary systems with a total of 22 detected exoplanets. Their orbit radii computed by A[n] exhibit an average deviation $\sim 4\%$ from the corresponding observed distances. In the systems examined, there are 50 unoccupied internal shells; excluding the 26 internal shells of the $\mu$~Ara system, this total drops to 24. These 24 internal shells, each providing 3 hosting orbits, can accomodate a total of $24 \times 3 = 72$ exoplanets. We have compared 30 of our predictions with respective ones given in Table~2 of \citep{BL13}, estimated by the generalized Titius--Bode relation, and we have found an average discrepancy $\sim 6\%$. In 5 cases, we have applied a conjecture asserting that, eventually, hosting orbits are average-density orbits instead of respective maximum-density orbits, provided that the latter should be already occupied. Finally, in 3 cases with large discrepancies $(\geq 15\%)$, we have adopted our predictions in the place of the corresponding predictions of \citep{BL13}.

\begin{table}
\begin{center}
\caption{The HD~40307 system: central body $S_1$, i.e. the host star HD~40307, and polytropic spherical shells of the planets b, c, d, e, f, g. For successive shells $S_j$ and $S_{j+1}$, inner radius of $S_{j+1}$ is the outer radius of $S_j$. All radii are expressed in AU, except for the host's radius $\xi_1$. Percent errors $\%E_j$ in the computed orbit radii $\alpha_j$ are given with respect to the corresponding observed radii $A_j$, $\%E_j = 100 \times |(A_j - \alpha_j)| / A_j$. Parenthesized signed integers following numerical values denote powers of 10. \label{hd40307}}
\begin{tabular}{lrrl} 
\hline \hline
Host star HD~40307 -- Shell No                  & 1              \\
$n_\mathrm{opt}$                                & $2.531\,\,\,(+00)$  \\
$\xi_1$ (cpu)                                   & $5.4317(+00)$  \\
$\xi_1$ ($R_\odot$)                             & $7.16 \ \ \ (-01)$  \\
\hline
            & & $A$~~~~~~ & $~~~~\%E$ \\  
\hline

b -- Shell No                                 & 5                 \\
Inner radius, $\, \xi_4$                      & $3.7372(-02)$     \\              
Outer radius, $\xi_5$                         & $6.4264(-02)$     \\
Orbit radius, $\, \alpha_\mathrm{b}=\alpha_5$ & $4.8999(-02)$ & $4.68(-02)$ & $4.70(+00)$ \\
\hline

c -- Shell No                                 & 6                 \\
Outer radius, $\xi_6$                         & $9.4481(-02)$     \\
Orbit radius, $\, \alpha_\mathrm{c}=\alpha_6$ & $8.2084(-02)$ & $7.99(-02)$ & $2.73(+00)$ \\
\hline

d -- Shell No                                 & 8                 \\
Inner radius, $\, \xi_7$                      & $1.2142(-01)$     \\
Outer radius, $\xi_8$                         & $1.6372(-01)$     \\
Orbit radius, $\, \alpha_\mathrm{d}=\alpha_8$ & $1.3639(-01)$ & $1.321(-01)$ & $3.25(+00)$ \\   
\hline

e -- Shell No                                 & 9                 \\
Outer radius, $\xi_9$                         & $2.0736(-01)$     \\
Orbit radius, $\, \alpha_\mathrm{e}=\alpha_9$ & $2.0636(-01)$ & $1.886(-01)$ & $9.42(+00)$               \\ 
\hline

f -- Shell No                                    & 10                \\
Outer radius, $\xi_{10}$                         & $2.5487(-01)$     \\
Orbit radius, $\, \alpha_\mathrm{f}=\alpha_{10}$ & $2.0838(-01)$ & $2.47(-01)$ & $1.56(+01)$               \\   
\hline

g -- Shell No                                    & 15                \\
Inner radius, $\, \xi_{14}$                      & $5.3513(-01)$     \\
Outer radius, $\xi_{15}$                         & $6.3922(-01)$     \\
Orbit radius, $\, \alpha_\mathrm{g}=\alpha_{15}$ & $5.8682(-01)$ & $6.(-01)$ & $2.20(+00)$               \\   
\hline

\end{tabular}
\end{center}
\end{table}

\begin{table}
\begin{center}
\caption{The HD~40307 system: unoccupied shells, or unoccupied orbits $\alpha_\mathrm{Lj}$,  $\alpha_j$, $\alpha_\mathrm{Rj}$ within occupied shells. $P_j$ denote predicted orbit radii of planets not yet observed, according to Table~2 of \citep{BL13}; external shells, i.e. shells next to the outermost occupied shell, are included in this table only when such predictions are available. Percent differences $\%D_j$ in the computed orbit radii $\alpha_j$ are given with respect to the corresponding predicted radii $P_j$, $\%D_j = 100 \times |(P_j - \alpha_j)| / P_j$. Other details as in Table \ref{hd40307}.\label{hd40307IE}}
\begin{tabular}{lrrl} 
\hline \hline
            & & $P$~~~~ & $~~~~\%D$ \\  
\hline  
Shell No                                & 2  &  &           \\
Inner radius, $\, \xi_1$                & $3.3310(-03)$     \\
Outer radius, $\xi_2$                   & $1.1002(-02)$     \\ 
Orbit radius, $\, \alpha_2$             & $7.0832(-03)$     \\
\hline

Shell No                                & 3                 \\
Outer radius, $\xi_3$                   & $2.0234(-02)$     \\ 
Orbit radius, $\, \alpha_3$             & $1.4473(-02)$     \\
\hline

Shell No                                & 4                 \\
Outer radius, $\xi_4$                   & $3.7372(-02)$     \\ 
Orbit radius, $\, \alpha_4$             & $2.5603(-02)$     \\
\hline

Shell No                                & 5                 \\
Outer radius, $\xi_5$                   & $6.4264(-02)$     \\
Orbit radius, $\, \alpha_\mathrm{R5}$   & $5.7002(-02)$  &  $6.(-02)$  & $5.00(+00)$ \\   
\hline

Shell No                                & 7                 \\
Inner radius, $\, \xi_6$                & $9.4481(-02)$     \\
Outer radius, $\xi_7$                   & $1.2142(-01)$     \\
Orbit radius, $\, \alpha_7$             & $1.0609(-01)$  &  $1.1(-01)$  & $3.55(+00)$  \\
\hline

Shell No                                & 11                \\
Inner radius, $\, \xi_{10}$             & $2.5487(-01)$     \\
Outer radius, $\xi_{11}$                & $3.2477(-01)$     \\
Orbit radius, $\, \alpha_{11}$          & $2.8608(-01)$     \\
\hline

Shell No                                & 12                \\
Outer radius, $\xi_{12}$                & $3.9435(-01)$     \\
Orbit radius, $\, \alpha_{12}$          & $3.6973(-01)$  &  $3.3(-01)$ & $1.20(+01)$   \\

\hline

Shell No                                & 13                \\
Outer radius, $\xi_{13}$                & $4.4884(-01)$     \\
Orbit radius, $\, \alpha_{13}$          & $4.1694(-01)$  &  $4.4(-01)$  & $5.24(+00)$   \\
\hline

Shell No                                & 14                \\
Outer radius, $\xi_{14}$                & $5.3513(-01)$     \\
Orbit radius, $\, \alpha_{14}$          & $4.8061(-01)$  &  \\
\hline

External Shell No                       & 17                \\
Inner radius, $\, \xi_{16}$             & $7.2715(-01)$     \\
Outer radius, $\xi_{17}$                & $8.2144(-01)$     \\
Orbit radius, $\, \alpha_{17}$          & $7.3545(-01)$  &  $7.8(-01)$  & $5.71(+00)$   \\
\hline

\end{tabular}
\end{center}
\end{table}

\begin{table}
\begin{center}
\caption{The $\mu$~Ara system: central body $S_1$, i.e. the host star $\mu$~Ara, and polytropic spherical shells of the planets c, d, b, e. Other details as in Table~\ref{hd40307}. \label{mAr}}
\begin{tabular}{lrrl} 
\hline \hline
Host star $\mu$~Ara -- Shell No                    & 1              \\
$n_\mathrm{opt}$                                & $2.554\,\,\,(+00)$  \\
$\xi_1$ (cpu)                                   & $5.4898(+00)$  \\
$\xi_1$ ($R_\odot$)                             & $1.245\,\,\,(+00)$  \\
\hline
            & & $A$~~~~~~ & $~~~~\%E$ \\  
\hline

c -- Shell No                                 & 5                 \\
Inner radius, $\, \xi_4$                      & $6.6323(-02)$     \\              
Outer radius, $\xi_5$                         & $1.1495(-01)$     \\
Orbit radius, $\, \alpha_\mathrm{c}=\alpha_5$ & $8.7360(-02)$ & $9.094(-02)$ & $3.94(+00)$ \\
\hline

d -- Shell No                                 & 14                 \\
Inner radius, $\, \xi_{13}$                   & $8.2970(-01)$     \\
Outer radius, $\xi_{14}$                      & $9.7484(-01)$     \\
Orbit radius, $\, \alpha_\mathrm{d}=\alpha_{14}$ & $9.1279(-01)$ & $9.210(-01)$ & $8.91(-01)$ \\   
\hline

b -- Shell No                                 & 18                 \\
Inner radius, $\, \xi_{17}$                   & $1.4513(+00)$      \\
Outer radius, $\xi_{18}$                      & $1.6443(+00)$      \\
Orbit radius, $\, \alpha_\mathrm{b}=\alpha_{18}$ & $1.5075(+00)$ & $1.5(+00)$ & $5.03(-01)$               \\ 
\hline

e -- Shell No                                    & 31                \\
Inner radius, $\, \xi_{30}$                      & $5.0053(+00)$     \\
Outer radius, $\xi_{31}$                         & $5.4708(+00)$     \\
Orbit radius, $\, \alpha_\mathrm{e}=\alpha_{31}$ & $5.2381(+00)$ & $5.235(+00)$ & $5.84(-02)$               \\   
\hline

\end{tabular}
\end{center}
\end{table}

\begin{table}
\begin{center}
\caption{The $\mu$~Ara system: unoccupied shells for which predicted orbit radii $P_j$ are given for planets not yet observed, according to Table~2 of \citep{BL13}. Other details as in Table \ref{hd40307IE}. \label{mArIE}}
\begin{tabular}{lrrl} 
\hline \hline
            & & $P$~~~~ & $~~~~\%D$ \\  
\hline  
Shell No                                & 6  &  &           \\
Inner radius, $\, \xi_5$                & $1.1495(-01)$     \\
Outer radius, $\xi_6$                   & $1.6901(-01)$     \\ 
Orbit radius, $\, \alpha_6$             & $1.4882(-01)$ & $1.6(-01)$ & $6.99(+00)$     \\
\hline

Shell No                                & 8                 \\
Inner radius, $\, \xi_7$                & $2.1725(-01)$     \\
Outer radius, $\xi_8$                   & $2.9575(-01)$     \\ 
Orbit radius, $\, \alpha_8$             & $2.4604(-01)$ & $2.9(-01)$ & $1.52(+01)$     \\
\hline

Shell No                                & 11                \\
Inner radius, $\, \xi_{10}$             & $4.6303(-01)$     \\
Outer radius, $\xi_{11}$                & $5.9148(-01)$     \\
Orbit radius, $\, \alpha_{11}$          & $5.2105(-01)$ & $5.2(-01)$ & $2.02(-01)$     \\
\hline

Shell No                                & 24                \\
Inner radius, $\, \xi_{23}$             & $2.8301(+00)$     \\
Outer radius, $\xi_{24}$                & $3.1098(+00)$     \\
Orbit radius, $\, \alpha_{24}$          & $3.0123(+00)$ & $2.9(+00)$ & $3.87(+00)$     \\
\hline

External Shell No                       & 40                \\
Inner radius, $\, \xi_{39}$             & $8.9572(+00)$     \\
Outer radius, $\xi_{40}$                & $9.4795(+00)$     \\
Orbit radius, $\, \alpha_{40}$          & $9.1042(+00)$ &  $9.17(+00)$  & $7.18(-01)$   \\
\hline

\end{tabular}
\end{center}
\end{table}

\begin{table}
\begin{center}
\caption{The Kepler-26 system: central body $S_1$, i.e. the host star Kepler-26, and polytropic spherical shells of the planets d, b, c, e. Other details as in Table~\ref{hd40307}. \label{k26}}
\begin{tabular}{lrrl} 
\hline \hline
Host star Kepler-26 -- Shell No                 & 1              \\
$n_\mathrm{opt}$                                & $2.531\,\,\,(+00)$  \\
$\xi_1$ (cpu)                                   & $5.4317(+00)$  \\
$\xi_1$ ($R_\odot$)                             & $5.9 \ \ \ \ \, (-01)$  \\
\hline
            & & $A$~~~~~~ & $~~~~\%E$ \\  
\hline

d -- Shell No                                 & 5                 \\
Inner radius, $\, \xi_{4}$                    & $3.0795(-02)$     \\
Outer radius, $\xi_{5}$                       & $5.2955(-02)$     \\
Orbit radius, $\, \alpha_\mathrm{d}=\alpha_{5}$ & $4.0376(-02)$ & $3.9(-02)$ & $3.53(+00)$ \\   
\hline

b -- Shell No                                 & 7                 \\
Inner radius, $\, \xi_{6}$                    & $7.7855(-02)$     \\
Outer radius, $\xi_{7}$                       & $1.0005(-01)$     \\
Orbit radius, $\, \alpha_\mathrm{b}=\alpha_{7}$ & $8.7423(-02)$ & $8.5(-02)$ & $2.85(+00)$               \\ 
\hline

c -- Shell No                                 & 8                 \\
Outer radius, $\xi_8$                         & $1.3491(-01)$     \\
Orbit radius, $\, \alpha_\mathrm{c}=\alpha_8$ & $1.1239(-01)$ & $1.07(-01)$ & $5.03(+00)$ \\
\hline

e -- Shell No                                    & 11                \\
Inner radius, $\, \xi_{10}$                      & $2.1002(-01)$     \\
Outer radius, $\xi_{11}$                         & $2.6762(-01)$     \\
Orbit radius, $\, \alpha_\mathrm{e}=\alpha_{11}$ & $2.3574(-01)$ & $2.2(-01)$ & $7.15(+00)$               \\   
\hline

\end{tabular}
\end{center}
\end{table}

\begin{table}
\begin{center}
\caption{The Kepler-26 system: unoccupied shells, or unoccupied orbits $\alpha_\mathrm{Lj}$, $\alpha_j$, $\alpha_\mathrm{Rj}$ within occupied shells. $P_j$ denote predicted orbit radii of planets not yet observed, according to Table~2 of \citep{BL13}. Other details as in Table \ref{hd40307IE}. \label{k26IE}}
\begin{tabular}{lrrl} 
\hline \hline
            & & $P$~~~~ & $~~~~\%D$ \\  
\hline
Shell No                                & 2  &  &           \\
Inner radius, $\, \xi_1$                & $2.7448(-03)$     \\
Outer radius, $\xi_2$                   & $9.0659(-03)$     \\ 
Orbit radius, $\, \alpha_2$             & $5.8367(-03)$     \\
\hline

Shell No                                & 3                 \\
Outer radius, $\xi_3$                   & $1.6674(-02)$     \\ 
Orbit radius, $\, \alpha_3$             & $1.1926(-02)$     \\
\hline

Shell No                                & 4                 \\
Outer radius, $\xi_4$                   & $3.07954(-02)$     \\ 
Orbit radius, $\, \alpha_4$             & $2.10978(-02)$     \\
\hline

Shell No                                & 5                 \\
Outer radius, $\xi_5$                   & $5.2955(-02)$     \\
Orbit radius, $\, \alpha_\mathrm{R5}$   & $4.6971(-02)$ & $5.(-02)$ & $6.06(+00)$ \\
\hline

Shell No                                & 6                 \\
Outer radius, $\xi_6$                   & $7.7855(-02)$     \\
Orbit radius, $\, \alpha_\mathrm{L6}$   & $5.9882(-02)$ & $6.(-02)$ & $1.97(-01)$ \\
Orbit radius, $\, \alpha_6$             & $6.7639(-02)$ & $7.(-02)$ & $3.37(+00)$  \\
\hline

Shell No                                & 9                 \\
Inner radius, $\, \xi_8$                & $1.3491(-01)$     \\
Outer radius, $\xi_9$                   & $1.7087(-01)$     \\ 
Orbit radius, $\, \alpha_\mathrm{L9}$   & $1.6284(-01)$ & $1.4(-01)$ & $1.63(+01)$ \\
Orbit radius, $\, \alpha_9$             & $1.7005(-01)$ & $1.7(-01)$ & $2.94(-02)$     \\
\hline

Shell No                                & 10                \\
Outer radius, $\xi_{10}$                & $2.1002(-01)$     \\
Orbit radius, $\, \alpha_{10}$          & $1.7171(-01)$     \\
\hline

External Shell No                       & 12                \\
Inner radius, $\, \xi_{11}$             & $2.6762(-01)$     \\
Outer radius, $\xi_{12}$                & $3.2496(-01)$     \\
Orbit radius, $\, \alpha_{12}$          & $3.0467(-01)$ & $2.7(-01)$ & $1.28(+01)$   \\
\hline

\end{tabular}
\end{center}
\end{table}

\begin{table}
\begin{center}
\caption{The Kepler-62 system: central body $S_1$, i.e. the host star Kepler-62, and polytropic spherical shells of the planets b, c, d, e, f. Other details as in Table \ref{hd40307}. \label{k62}}
\begin{tabular}{lrrl} 
\hline \hline
Host star Kepler-62 -- Shell No                 & 1              \\
$n_\mathrm{opt}$                                & $2.773\,\,\,(+00)$  \\
$\xi_1$ (cpu)                                   & $6.1063(+00)$  \\
$\xi_1$ ($R_\odot$)                             & $6.3 \ \ \ \ \, (-01)$  \\
\hline
            & & $A$~~~~~~ & $~~~~\%E$ \\  
\hline

b -- Shell No                                 & 5                 \\
Inner radius, $\, \xi_4$                      & $4.3481(-02)$     \\              
Outer radius, $\xi_5$                         & $7.6678(-02)$     \\
Orbit radius, $\, \alpha_\mathrm{b}=\alpha_5$ & $5.5342(-02)$ & $5.53(-02)$ & $7.65(-02)$ \\
\hline

c -- Shell No                                 & 6                 \\
Outer radius, $\xi_6$                         & $1.1442(-01)$     \\
Orbit radius, $\, \alpha_\mathrm{c}=\alpha_6$ & $1.0545(-01)$ & $9.29(-02)$ & $1.35(+01)$ \\
\hline

d -- Shell No                                 & 7                 \\
Outer radius, $\xi_{7}$                       & $1.5633(-01)$     \\
Orbit radius, $\, \alpha_\mathrm{d}=\alpha_{7}$ & $1.2283(-01)$ & $1.2(-01)$ & $2.36(+00)$ \\   
\hline

e -- Shell No                                    & 11                \\
Inner radius, $\, \xi_{10}$                      & $3.5889(-01)$     \\
Outer radius, $\xi_{11}$                         & $4.4702(-01)$     \\
Orbit radius, $\, \alpha_\mathrm{e}=\alpha_{11}$ & $4.4511(-01)$ & $4.27(-01)$ & $4.24(-01)$               \\ 
\hline

f -- Shell No                                    & 14                \\
Inner radius, $\, \xi_{13}$                      & $6.7326(-01)$     \\
Outer radius, $\xi_{14}$                         & $7.8344(-01)$     \\
Orbit radius, $\, \alpha_\mathrm{f}=\alpha_{14}$ & $7.1679(-01)$ & $7.18(-01)$ & $1.69(-01)$               \\   
\hline

\end{tabular}
\end{center}
\end{table}

\begin{table}
\begin{center}
\caption{The Kepler-62 system: unoccupied shells, or unoccupied orbits $\alpha_\mathrm{Lj}$, $\alpha_j$, $\alpha_\mathrm{Rj}$ within occupied shells. $P_j$ denote predicted orbit radii of planets not yet observed, according to Table~2 of \citep{BL13}. Other details as in Table \ref{hd40307IE}. \label{k62IE}}
\begin{tabular}{lrrl} 
\hline \hline
            & & $P$~~~~ & $~~~~\%D$ \\  
\hline
Shell No                                & 2  &  &           \\
Inner radius, $\, \xi_1$                & $2.9309(-03)$     \\
Outer radius, $\xi_2$                   & $1.2620(-02)$     \\ 
Orbit radius, $\, \alpha_2$             & $6.2399(-03)$     \\
\hline

Shell No                                & 3                 \\
Outer radius, $\xi_3$                   & $2.6848(-02)$     \\ 
Orbit radius, $\, \alpha_3$             & $2.3207(-02)$     \\
\hline

Shell No                                & 4                 \\
Outer radius, $\xi_4$                   & $4.3481(-02)$     \\ 
Orbit radius, $\, \alpha_4$             & $3.0100(-02)$     \\
\hline

Shell No                                & 5                 \\
Outer radius, $\xi_5$                   & $7.6678(-02)$     \\
Orbit radius, $\, \alpha_\mathrm{R5}$   & $6.6406(-02)$ & $7.(-02)$ & $5.13(+00)$ \\
\hline

Shell No                                & 7                 \\
Inner radius, $\, \xi_6$                & $1.1442(-01)$     \\
Outer radius, $\xi_7$                   & $1.5633(-01)$     \\
Orbit radius, $\, \alpha_\mathrm{R7}$   & $1.4004(-01)$ & $1.5(-01)$ & $6.64(+00)$ \\
\hline

Shell No                                & 8                 \\
Outer radius, $\xi_8$                   & $2.1787(-01)$     \\ 
Orbit radius, $\, \alpha_\mathrm{8 }$   & $1.8604(-01)$ & $2.(-01)$ & $6.98(+00)$ \\
\hline

Shell No                                & 9                 \\
Outer radius, $\xi_{9}$                 & $2.7712(-01)$     \\
Orbit radius, $\, \alpha_{9}$           & $2.5403(-01)$ & $2.6(-01)$ & $2.30(+00)$    \\
\hline

Shell No                                 & 10                 \\
Outer radius, $\xi_{10}$                 & $3.5889(-01)$      \\
Orbit radius, $\, \alpha_{10}$           & $2.9922(-01)$      \\
Orbit radius, $\, \alpha_\mathrm{R10}$   & $3.2838(-01)$ & $3.3(-01)$ & $4.90(-01)$ \\
\hline

Shell No                                 & 12                 \\
Inner radius, $\, \xi_{11}$              & $4.4702(-01)$     \\
Outer radius, $\xi_{12}$                 & $5.4706(-01)$      \\
Orbit radius, $\, \alpha_{12}$           & $4.4893(-01)$      \\
\hline

Shell No                                 & 13                 \\
Outer radius, $\xi_{13}$                 & $6.7326(-01)$      \\
Orbit radius, $\, \alpha_{13}$           & $6.2825(-01)$ & $5.5(-01)$ & $1.42(+01)$      \\
\hline

External Shell No                        & 15                \\
Inner radius, $\, \xi_{14}$              & $7.8344(-01)$     \\
Outer radius, $\xi_{15}$                 & $9.4463(-01)$     \\
Orbit radius, $\, \alpha_{15}$           & $8.4677(-01)$  &  $9.2(-01)$  & $7.96(+00)$   \\
\hline

\end{tabular}
\end{center}
\end{table}

\begin{table}
\begin{center}
\caption{The Kepler-275 system: central body $S_1$, i.e. the host star Kepler-275, and polytropic spherical shells of the planets b, c, d. Other details as in Table~\ref{hd40307}. \label{k275}}
\begin{tabular}{lrrl} 
\hline \hline
Host star Kepler-275 -- Shell No                & 1              \\
$n_\mathrm{opt}$                                & $2.837\,\,\,(+00)$  \\
$\xi_1$ (cpu)                                   & $6.3115(+00)$  \\
$\xi_1$ ($R_\odot$)                             & $1.38 \ \ \ (+00)$  \\
\hline
            & & $A$~~~~~~ & $~~~~\%E$ \\  
\hline

b -- Shell No                                 & 4                 \\
Inner radius, $\, \xi_{3}$                    & $7.1357(-02)$     \\
Outer radius, $\xi_{4}$                       & $1.1188(-01)$     \\
Orbit radius, $\, \alpha_\mathrm{b}=\alpha_{4}$ & $9.8416(-02)$ & $9.8(-02)$ & $4.24(-01)$ \\   
\hline

c -- Shell No                                 & 5                 \\
Outer radius, $\xi_{5}$                       & $1.8278(-01)$     \\
Orbit radius, $\, \alpha_\mathrm{c}=\alpha_{5}$ & $1.2481(-01)$ & $1.32(-01)$ & $5.44(+00)$               \\ 
\hline

d -- Shell No                                    & 6                \\
Outer radius, $\xi_{6}$                          & $3.0822(-01)$     \\
Orbit radius, $\, \alpha_\mathrm{d}=\alpha_{6}$ & $2.3432(-01)$ & $2.24(-01)$ & $4.61(+00)$               \\   
\hline

\end{tabular}
\end{center}
\end{table}

\begin{table}
\begin{center}
\caption{The Kepler-275 system: unoccupied shells, or unoccupied orbits $\alpha_\mathrm{Lj}$, $\alpha_j$, $\alpha_\mathrm{Rj}$ within occupied shells. $P_j$ denote predicted orbit radii of planets not yet observed, according to Table~2 of \citep{BL13}. Other details as in Table \ref{hd40307IE}. \label{k275IE}}
\begin{tabular}{lrrl} 
\hline \hline
            & & $P$~~~~ & $~~~~\%D$ \\  
\hline
Shell No                                & 2  &  &           \\
Inner radius, $\, \xi_1$                & $6.4201(-03)$     \\
Outer radius, $\xi_2$                   & $2.9258(-02)$     \\
Orbit radius, $\, \alpha_2$             & $1.3834(-02)$     \\ 
Orbit radius, $\, \alpha_\mathrm{R2}$   & $2.2430(-02)$ & $3.(-02)$ & $2.52(+01)$ \\
\hline

Shell No                                & 3                 \\
Outer radius, $\xi_3$                   & $7.1357(-02)$     \\
Orbit radius, $\, \alpha_\mathrm{L3}$   & $3.8197(-02)$ & $4.(-02)$ & $4.51(+00)$ \\ 
Orbit radius, $\, \alpha_3$             & $5.1166(-02)$ & $5.(-02)$ & $2.33(+00)$ \\
Orbit radius, $\, \alpha_\mathrm{R3}$   & $6.0524(-02)$ & $6.(-02)$ & $8.73(-01)$ \\
\hline

Shell No                                & 4                 \\
Outer radius, $\xi_{4}$                 & $1.1188(-01)$     \\
Orbit radius, $\, \alpha_\mathrm{L4}$   & $8.2954(-02)$ & $8.(-02)$ & $3.69(+00)$ \\
\hline

Shell No                                & 5                 \\
Outer radius, $\xi_5$                   & $1.8278(-01)$     \\
Orbit radius, $\, \alpha_\mathrm{R5}$   & $1.4480(-01)$ & $1.7(-01)$ & $1.48(+01)$ \\
\hline

Shell No                                & 6                 \\
Outer radius, $\xi_6$                   & $3.0822(-01)$     \\
Orbit radius, $\, \alpha_\mathrm{R6}$   & $2.6992(-01)$ & $2.9(-01)$ & $6.92(+00)$ \\
\hline

\end{tabular}
\end{center}
\end{table}

\end{document}